\documentclass{webofc}
\usepackage[varg]{txfonts}   

\begin{document}

\title{The d$^\ast$(2380) and its family of pion assisted dibaryons} 

\author{\firstname{Avraham} \lastname{Gal}\inst{1}\fnsep
\thanks{updated version of talks given at MESON 2018, Krak\'{o}w 
\cite{Gal19a}, and at QNP 2018, Tsukuba \cite{Gal19b}} 
\institute{Racah Institute of Physics, the Hebrew University, 
Jerusalem 91904, Israel (avragal@savion.huji.ac.il)}}

\abstract
{The status of $N\Delta$ and $\Delta\Delta$ dibaryons introduced by Dyson 
and Xuong in 1964 is briefly reviewed with focus on the d$^\ast$(2380), 
tentatively assigned as a $\Delta\Delta$ dibaryon resonance. It is argued 
that the apparently small value of width, $\Gamma_{d^\ast}$$\approx$70~MeV, 
favors hadronic structure for the d$^\ast$(2380) dibaryon rather than 
a six-quark structure.}  
\maketitle

\section{Pion assisted $N\Delta$ and $\Delta\Delta$ dibaryons} 
\label{sec:int} 

Nonstrange $s$-wave dibaryon resonances ${\cal D}_{IS}$ with isospin $I$ 
and spin $S$ were predicted by Dyson and Xuong in 1964~\cite{DX64} as early 
as SU(6) symmetry proved successful, placing the nucleon $N(939)$ and its 
$P_{33}$ $\pi N$ resonance $\Delta(1232)$ in the same ${\bf 56}$ multiplet 
which reduces to a ${\bf 20}$ SU(4) spin-isospin multiplet for nonstrange 
baryons. For SU(3)-color singlet and spatially symmetric $L=0$ 6q 
configuration, the spin-isospin 6q configuration ensuring a totally 
antisymmetric color-spin-isospin-space 6q wavefunction is a ${\bf 50}$ 
dimensional SU(4) representation, denoted by its (3,3,0,0) Young tableau, 
which is the lowest-dimension SU(4) multiplet in the $\bf{20\times 20}$ 
direct product~\cite{PPL15}. This ${\bf 50}$ SU(4) multiplet includes the 
deuteron ${\cal D}_{01}$ and $NN$ virtual state ${\cal D}_{10}$, plus four 
more nonstrange dibaryons, with masses listed in Table~\ref{tab:dyson} in 
terms of SU(4) mass-formula constants $A$ and $B$. 

\begin{table}[hbt] 
\begin{center}
\caption{Predicted masses of non-strange $L=0$ dibaryons ${\cal D}_{IS}$ 
with isospin $I$ and spin $S$, using the Dyson-Xuong~\cite{DX64} 
SU(6)$\to$SU(4) mass formula $M=A+B\,[I(I+1)+S(S+1)-2]$.} 
\begin{tabular}{ccccccccccccc}
\hline
${\cal D}_{IS}$ & & ${\cal D}_{01}$ & & ${\cal D}_{10}$ & & ${\cal D}_{12}$ 
& & ${\cal D}_{21}$ & & ${\cal D}_{03}$ & & ${\cal D}_{30}$ \\
\hline
$BB'$ & & $NN$ & & $NN$ & & $N\Delta$ & & $N\Delta$ & & $\Delta\Delta$ & & 
$\Delta\Delta$ \\
SU(3)$_{\rm f}$ & & $\overline{\bf 10}$ & & ${\bf 27}$ & & ${\bf 27}$ & & 
${\bf 35}$ & & $\overline{\bf 10}$ & & ${\bf 28}$ \\
$M({\cal D}_{IS})$ & & $A$ & & $A$ & & $A+6B$ & & $A+6B$ & & $A+10B$ & & 
$A+10B$ \\
\hline
\end{tabular}
\label{tab:dyson}
\end{center}
\end{table} 

Identifying $A$ with the $NN$ threshold mass 1878~MeV, the value 
$B\approx 47$~MeV was derived by assigning ${\cal D}_{12}$ to the 
$pp\leftrightarrow \pi^+ d$ coupled-channel resonance behavior noted then 
at 2160~MeV, near the $N\Delta$ threshold (2.171~MeV). This led in particular 
to a predicted mass $M=2350$~MeV for the $\Delta\Delta$ dibaryon candidate 
${\cal D}_{03}$ assigned at present to the recently established d$^\ast$(2380) 
resonance~\cite{clement17}. Since the ${\bf 27}$ and $\overline{\bf 10}$ 
flavor-SU(3) multiplets accommodate $NN$ $s$-wave states that are close 
to binding ($^1S_0$) or weakly bound ($^3S_1$), we focus here on the 
${\cal D}_{12}$ and ${\cal D}_{03}$ dibaryon candidates assigned to these 
flavor-SU(3) multiplets. 

The idea behind the concept of pion assisted dibaryons~\cite{gal16} is that 
since the $\pi N$ $p$-wave interaction in the $P_{33}$ channel is so strong 
as to form the $\Delta$(1232) baryon resonance, acting on two nucleons it may 
assist in forming $s$-wave $N\Delta$ dibaryon states, and subsequently also 
in forming $s$-wave $\Delta\Delta$ dibaryon states. This goes beyond the 
major role played by a $t$-channel exchange low-mass pion in binding or 
almost binding $NN$ $s$-wave states. 

As discussed below, describing $N\Delta$ systems in terms of 
a stable nucleon ($N$) and a two-body $\pi N$ resonance ($\Delta$) leads to 
a well defined $\pi NN$ three-body model in which $IJ=12$ and $21$ resonances 
identified with the ${\cal D}_{12}$ and ${\cal D}_{21}$ dibaryons of 
Table~\ref{tab:dyson} are generated. This relationship between $N\Delta$ and 
$\pi NN$ may be generalized into relationship between a two-body $B\Delta$ 
system and a three-body $\pi NB$ system, where the baryon $B$ stands for $N, 
\Delta, Y$ (hyperon) etc. In order to stay within a three-body formulation 
one needs to assume that the baryon $B$ is stable. For $B=N$, this 
formulation relates the $N\Delta$ system to the three-body $\pi NN$ system. 
For $B=\Delta$, once properly formulated, it relates the 
$\Delta\Delta$ system to the three-body $\pi N\Delta$ system, suggesting to 
seek $\Delta\Delta$ dibaryon resonances by solving $\pi N\Delta$ Faddeev 
equations, with a stable $\Delta$. The decay width of the $\Delta$ resonance 
is considered then at the penultimate stage of the calculation. In terms 
of two-body isobars we have then a coupled-channel problem $B\Delta
\leftrightarrow\pi D$, where $D$ stands generically for appropriate dibaryon 
isobars: (i) ${\cal D}_{01}$ and ${\cal D}_{10}$, which are the $NN$ isobars 
identified with the deuteron and virtual state respectively, for $B=N$, and 
(ii) ${\cal D}_{12}$ and ${\cal D}_{21}$ for $B=\Delta$. 

\begin{figure}[hbt] 
\begin{center} 
\includegraphics[width=0.8\textwidth]{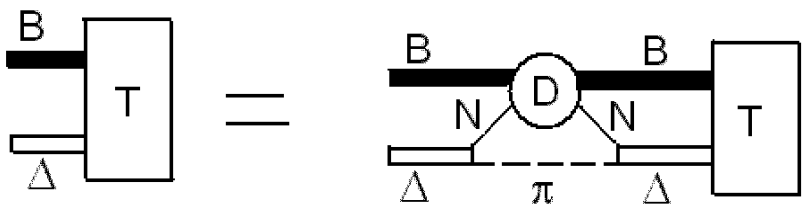} 
\caption{Diagrammatic representation of the $B\Delta$ $T$-matrix integral 
equation from $\pi NB$ Faddeev equations with separable pairwise 
interactions where $B=N,\,\Delta$~\cite{GG13,GG14}.} 
\label{fig:piNB} 
\end{center} 
\end{figure} 

Within this model, and using separable pairwise interactions, 
the coupled-channel $B\Delta -\pi D$ eigenvalue problem reduces to a single 
integral equation for the $B\Delta$ $T$ matrix shown diagrammatically in 
Fig.~\ref{fig:piNB}, where starting with a $B\Delta$ configuration the 
$\Delta$-resonance isobar decays into $\pi N$, followed by $NB\to NB$ 
scattering through the $D$-isobar with a spectator pion, and ultimately 
by means of the inverse decay $\pi N\to\Delta$ back into the $B\Delta$ 
configuration. The interaction between the $\pi$ meson and $B$ is neglected 
for $B=\Delta$, for lack of known $\pi\Delta$ isobar resonances in the 
relevant energy range. 

The ${\cal D}_{12}$ dibaryon of Table~\ref{tab:dyson} shows up clearly in 
the Argand diagram of the $NN$ $^1D_2$ partial wave which is coupled above 
the $NN\pi$ threshold to the $I=1$ $s$-wave $N\Delta$ channel. Its $S$-matrix 
pole position $W=M-{\rm i}\Gamma/2$ was given by 2148$-{\rm i}$63~MeV in $NN$ 
phase shift analyses~\cite{arndt87} and by 2144$-{\rm i}$55~MeV in dedicated 
$pp \leftrightarrow np\pi^+$ coupled-channels analyses~\cite{hosh92}. It has 
been observed, most likely, at $W$=(2.14$\pm$0.01)-i(0.09$\pm$0.01) GeV in 
a recent $\gamma d\to \pi^0\pi^0 d$ ELPH experiment~\cite{ELPH19}. Values of 
${\cal D}_{12}$ and ${\cal D}_{21}$ pole positions from our hadronic-model 
three-body $\pi NN$ Faddeev calculations~\cite{GG13,GG14}, substituting 
$N$ for $B$ in Fig.~\ref{fig:piNB}, are listed in Table~\ref{tab:BDel}. 
The ${\cal D}_{12}$ mass and width values calculated in the Faddeev hadronic 
model version using $r_{\Delta}\approx\,1.3$~fm are remarkably close to these 
phenomenologically derived values. As for the ${\cal D}_{21}$ dibaryon, recent 
$pp\to pp\pi^+\pi^-$ production data~\cite{wasa18} place it almost degenerate 
with the ${\cal D}_{12}$. Our $\pi NN$ Faddeev calculations produce it about 
10-20~MeV higher than the ${\cal D}_{12}$, see Table~\ref{tab:BDel}. The 
widths of these near-threshold $N\Delta$ dibaryons are, naturally, close to 
that of the $\Delta$ resonance. We note that only $^3S_1$ $NN$ enters the 
calculation of the ${\cal D}_{12}$ resonance, while for the ${\cal D}_{21}$ 
resonance calculation only $^1S_0$ $NN$ enters, both with maximal strength. 
Obviously, with the $^1S_0$ interaction the weaker of the two, one expects 
indeed that the ${\cal D}_{21}$ resonance lies above the ${\cal D}_{12}$ 
resonance. Moreover, these two dibaryon resonances differ also in their 
flavor-SU(3) classification, see Table~\ref{tab:dyson}, which is likely to 
push up the ${\cal D}_{21}$ further away from the ${\cal D}_{12}$. Finally, 
the $N\Delta$ $s$-wave states with $IJ=$ $11$ and $22$ are found not to 
resonate in the $\pi NN$ Faddeev calculations~\cite{GG14}. 

\begin{table}[hbt] 
\begin{center} 
\caption{${\cal D}_{IS}$ dibaryon $S$-matrix pole positions 
$M-{\rm i}\frac{\Gamma}{2}$ (in MeV) obtained by solving the $N\Delta$ 
and $\Delta\Delta$ $T$-matrix integral equation Fig.~\ref{fig:piNB} are 
listed for $\pi N$ $P_{33}$ form factors specified by radius parameter 
$r_{\Delta}$~\cite{GG13,GG14}.} 
\begin{tabular}{ccccc} 
\hline 
$r_{\Delta}$ & \multicolumn{2}{c}{$N\Delta$} & \multicolumn{2}{c}
{$\Delta\Delta$} \\ 
(fm) & ${\cal D}_{12}$ & ${\cal D}_{21}$ & ${\cal D}_{03}$ & ${\cal D}_{30}$ 
\\  
\hline 
1.3 & 2147$-{\rm i}$60 & 2165$-{\rm i}$64 & 2383$-{\rm i}$41 & 
2411$-{\rm i}$41 \\ 
0.9 & 2159$-{\rm i}$70 & 2169$-{\rm i}$69 & 2343$-{\rm i}$24 & 
2370$-{\rm i}$22 \\ 
\hline 
\end{tabular} 
\label{tab:BDel} 
\end{center} 
\end{table} 

The ${\cal D}_{03}$ dibaryon of Table~\ref{tab:dyson} is best demonstrated 
by the relatively narrow peak observed in $pn\to d\pi^0\pi^0$ by the 
WASA-at-COSY Collaboration~\cite{wasa11} about 80~MeV above the $\pi^0\pi^0$ 
production threshold and 80~MeV below the $\Delta\Delta$ threshold, with 
$\Gamma_{d^{\ast}}\approx 70$~MeV. Its $I=0$ isospin assignment follows from 
the isospin balance in $pn \to d\pi^0\pi^0$, and the $J^P=3^+$ spin-parity 
assignment follows from the measured deuteron angular distribution. The 
d$^{\ast}$(2380) was also observed in $pn\to d\pi^+\pi^-$~\cite{wasa13}, 
with cross section consistent with that measured in $pn\to d\pi^0\pi^0$, 
and studied in several $pn\to NN\pi\pi$ reactions~\cite{wasa15}. Recent 
measurements of $pn$ scattering and analyzing power~\cite{wasa14} have 
led to a $pn$ $^3D_3$ partial-wave Argand plot fully supporting the 
${\cal D}_{03}$ dibaryon resonance interpretation. 

Values of ${\cal D}_{03}$ and ${\cal D}_{30}$ pole positions $W=M-{\rm i}
\Gamma/2$ from our hadronic-model three-body $\pi N\Delta$ Faddeev 
calculations~\cite{GG13,GG14} are also listed in Table~\ref{tab:BDel}. 
The ${\cal D}_{03}$ mass and width values calculated in the Faddeev hadronic 
model version using $r_{\Delta}\approx\,1.3$~fm are remarkably close to the 
experimentally determined ones. The ${\cal D}_{30}$ dibaryon resonance is 
found in our $\pi N\Delta$ Faddeev calculations to lie about 30~MeV above 
the ${\cal D}_{03}$. These two states are degenerate in the limit of 
equal $D={\cal D}_{12}$ and $D={\cal D}_{21}$ isobar propagators in 
Fig.~\ref{fig:piNB}. Since ${\cal D}_{12}$ was found to lie lower than 
${\cal D}_{21}$, we expect also ${\cal D}_{03}$ to lie lower than 
${\cal D}_{30}$ as satisfied in our Faddeev calculations. Moreover, here 
too the difference in their flavor-SU(3) classification will push the 
${\cal D}_{30}$ further apart from the ${\cal D}_{03}$. The ${\cal D}_{30}$ 
has not been observed and only upper limits for its production in $pp\to pp
\pi^+\pi^+\pi^-\pi^-$ are available~\cite{wasa16}. 

Finally, we briefly discuss the ${\cal D}_{03}$ mass and width 
values from two recent quark-based resonating-group-method (RGM) 
calculations~\cite{wang14,dong16} that add $\Delta_{\bf 8}\Delta_{\bf 8}$ 
hidden-color (CC) components to a $\Delta_{\bf 1}\Delta_{\bf 1}$ cluster. 
The two listed calculations generate mass values 
that are close to the mass of the d$^{\ast}$(2380). The calculated widths, 
however, differ a lot from each other: one calculation generates a width of 
150~MeV~\cite{wang14}, exceeding substantially the reported value $\Gamma_{
d^{\ast}(2380)}$=80$\pm$10~MeV~\cite{wasa14}, the other one generates a width 
of 72~MeV~\cite{dong16}, thereby reproducing the d$^{\ast}$(2380) width. 
While the introduction of CC components has moderate effect on the resulting 
mass and width in the chiral version of the first calculation, lowering the 
mass by 20~MeV and the width by 25~MeV, it leads to substantial reduction of 
the width in the second (also chiral) calculation from 133~MeV to 72~MeV. 
The reason is that the dominant CC $\Delta_{\bf 8}\Delta_{\bf 8}$ components, 
with $68\%$ weight~\cite{dong16}, cannot decay through single-fermion 
transitions $\Delta_{\bf 8}\to N_{\bf 1}\pi_{\bf 1}$ to asymptotically 
free color-singlet hadrons. However, as argued in the next section, these 
quark-based width calculations miss important kinematical ingredients that 
make the width of a single compact $\Delta_{\bf 1}\Delta_{\bf 1}$ cluster 
considerably smaller than $\Gamma_{d^{\ast}(2380)}$. The introduction of 
substantial $\Delta_{\bf 8}\Delta_{\bf 8}$ components only aggravates the 
disagreement.

\section{The width of d$^\ast$(2380), small or large?} 
\label{sec:width} 

The width derived for the ${\cal D}_{03}$ dibaryon 
resonance d$^{\ast}$(2380) by WASA-at-COSY and SAID, 
$\Gamma_{d^{\ast}(2380)}$=80$\pm$10~MeV~\cite{wasa14}, is dominated by 
$\Gamma_{d^{\ast}\to NN\pi\pi}\approx 65$~MeV which is much smaller than 
twice the width $\Gamma_{\Delta}\approx 115$~MeV~\cite{SP07,anisovich12} of 
a single free-space $\Delta$, expected naively for a $\Delta\Delta$ quasibound 
configuration. However, considering the reduced phase space, $M_{\Delta}=1232 
\Rightarrow E_{\Delta}=1232-B_{\Delta\Delta}/2$~MeV in a bound-$\Delta$ decay, 
where $B_{\Delta\Delta}=2\times 1232-2380=84$~MeV is the $\Delta\Delta$ 
binding energy, the free-space $\Delta$ width gets reduced to 81~MeV using 
the in-medium single-$\Delta$ width $\Gamma_{\Delta\to N\pi}$ expression 
obtained from the empirical $\Delta$-decay momentum dependence 
\begin{equation}
\Gamma_{\Delta\to N\pi}(q_{\Delta\to N\pi})=\gamma\,
\frac{q^3_{\Delta\to N\pi}}{q_0^2+q^2_{\Delta\to N\pi}},
\label{eq:gamma}
\end{equation}
with $\gamma=0.74$ and $q_0=159$~MeV~\cite{BCS17}. Yet, this simple estimate 
is incomplete since neither of the two $\Delta$s is at rest in a deeply 
bound $\Delta\Delta$ state, as also noted by Niskanen~\cite{niskanen17}. 
To take account of the $\Delta\Delta$ momentum distribution, we evaluate 
the bound-$\Delta$ decay width ${\overline{\Gamma}}_{\Delta\to N\pi}$ by 
averaging $\Gamma_{\Delta\to N\pi}(\sqrt{s_{\Delta}})$ over the $\Delta\Delta$ 
bound-state momentum-space distribution~\cite{gal17}, 
\begin{equation}
{\overline{\Gamma}}_{\Delta\to N\pi}\equiv\langle \Psi^{\ast}(p_{\Delta\Delta})
|\Gamma_{\Delta\to N\pi}(\sqrt{s_{\Delta}})|\Psi(p_{\Delta\Delta})\rangle
\approx \Gamma_{\Delta\to N\pi}(\sqrt{{\overline{s}}_{\Delta}}), 
\label{eq:av}
\end{equation}
where $\Psi(p_{\Delta\Delta})$ is the $\Delta\Delta$ momentum-space 
wavefunction and the dependence of $\Gamma_{\Delta\to N\pi}$ on $q_{\Delta
\to N\pi}$ for on-mass-shell nucleons and pions was replaced by dependence 
on $\sqrt{s_{\Delta}}$. The averaged bound-$\Delta$ invariant energy squared 
${\overline{s}}_{\Delta}$ is defined by ${\overline{s}}_{\Delta}=(1232-B_{
\Delta\Delta}/2)^2-P_{\Delta\Delta}^2$ in terms of a $\Delta\Delta$ 
bound-state r.m.s. momentum $P_{\Delta\Delta}\equiv{\langle p_{\Delta
\Delta}^2\rangle}^{1/2}$ inversely proportional to the r.m.s. radius 
$R_{\Delta\Delta}$.  

The d$^{\ast}$(2380) in the quark-based RGM calculations of Ref.~\cite{dong16} 
appears quite squeezed compared to the diffuse deuteron. Its size, $R_{\Delta
\Delta}$=0.76~fm~\cite{huang15}, leads to unacceptably small upper limit of 
about 47~MeV for $\Gamma_{d^{\ast}\to NN\pi\pi}$~\cite{gal17}. This drastic 
effect of momentum dependence is missing in quark-based width calculations 
dealing with pionic decay modes of $\Delta_{\bf 1}\Delta_{\bf 1}$ components, 
e.g. Ref.~\cite{dong16}, and as presented by this Beijing group at MESON 2018 
\cite{huang19} and at QNP 2018 \cite{dong19a}. Practitioners of quark-based 
models ought therefore to ask ``what makes $\Gamma_{d^{\ast}(2380)}$ so much 
larger than the width calculated for a compact $\Delta\Delta$ dibaryon?" 
rather than ``what makes $\Gamma_{d^{\ast}(2380)}$ so much smaller than 
twice a free-space $\Delta$ width?" 

The preceding discussion of $\Gamma_{d^{\ast}(2380)}$ suggests that 
quark-based model findings of a tightly bound $\Delta\Delta$ $s$-wave 
configuration are in conflict with the observed width. Fortunately, 
hadronic-model calculations~\cite{GG13,GG14} offer resolution of this 
insufficiency by coupling to the tightly bound and compact $\Delta\Delta$ 
component of the d$^{\ast}$(2380) dibaryon's wavefunction a $\pi N\Delta$ 
resonating component dominated asymptotically by a $p$-wave pion attached 
loosely to the near-threshold $N\Delta$ dibaryon ${\cal D}_{12}$ with size 
about 1.5--2~fm. Formally, one can recouple spins and isospins in this 
$\pi{\cal D}_{12}$ system, so as to assume an extended $\Delta\Delta$-like 
object. This explains why a discussion of $\Gamma_{d^{\ast}\to NN\pi\pi}$ 
in terms of a $\Delta\Delta$ constituent model requires a size 
$R_{\Delta\Delta}$ considerably larger than provided by quark-based RGM 
calculations~\cite{dong16} to reconcile with the reported value of 
$\Gamma_{d^{\ast}(2380)}$. We recall that the width calculated in our 
diffuse-structure $\pi N\Delta$ model~\cite{GG13,GG14}, as listed in 
Table~\ref{tab:BDel}, is in good agreement with the observed width of 
the d$^{\ast}$(2380) dibaryon resonance. 

\begin{figure}[!t]
\begin{center}
\includegraphics[width=0.6\textwidth]{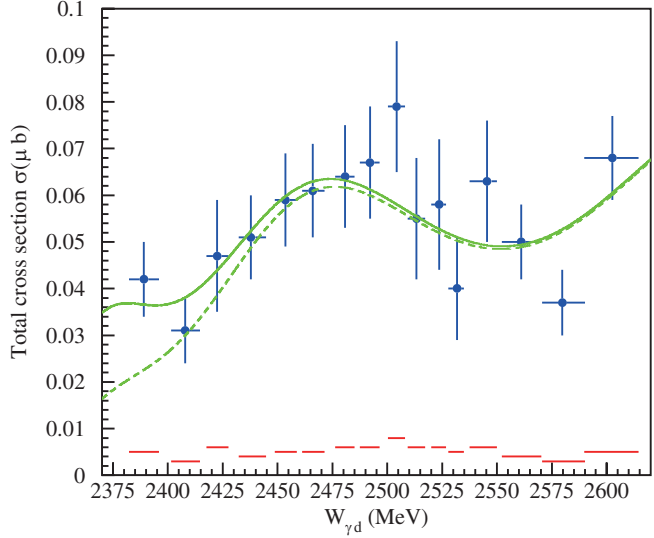} 
\caption{$\sigma(\gamma d\to d\pi^0\pi^0)$ as a function of the total cm 
energy $W$ from the ELPH experiment~\cite{ELPH17}. The red histogram shows 
systematic errors, the dotted curve shows a nonresonant calculation by Fix 
and Arenh\"{o}vel~\cite{fix05} and the solid curve is obtained by adding 
a BW shape centered at $W_{\gamma d}=2370$~MeV with $\Gamma=68$~MeV. 
A similar excitation spectrum has been reported by the MAMI A2 
Collaboration~\cite{MAMI18}.} 
\label{fig:fix} 
\end{center} 
\end{figure} 

\begin{figure}[!b]
\begin{center}
\includegraphics[width=0.7\textwidth]{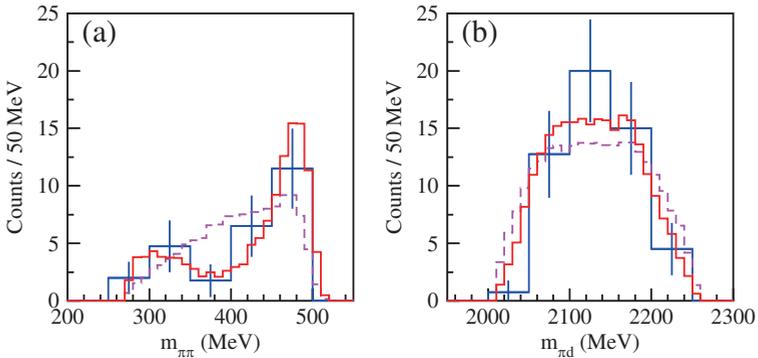} 
\caption{Invariant mass distributions in ELPH experiment~\cite{ELPH17} 
$\gamma d \to d \pi^0 \pi^0$ at $W_{\gamma d}=2.39$~GeV.} 
\label{fig:ELPH}
\end{center}               
\end{figure}

Support for the role of the $\pi{\cal D}_{12}$ configuration in the decay of 
the d$^{\ast}$(2380) dibaryon resonance is provided by a recent ELPH $\gamma d 
\to d \pi^0 \pi^0$ experiment~\cite{ELPH17}. The cross section data shown in 
Fig.~\ref{fig:fix} agree with a relativistic Breit-Wigner (BW) resonance shape 
centered at 2370~MeV and width of 68~MeV, but the statistical significance of 
the fit is low, particularly since most of the data are from the energy region 
above the d$^{\ast}$(2380). Invariant mass distributions from this experiment 
at $W_{\gamma d}=2.39$~GeV, shown in Fig.~\ref{fig:ELPH}, are more instructive. 
The $\pi\pi$ mass distribution shown in (a) suggests a two-bump structure, 
fitted in solid red. The lower bump around 300~MeV is perhaps a manifestation 
of the ABC effect~\cite{ABC60}, already observed in $pn\to d\pi^0\pi^0$ by 
WASA-at-COSY~\cite{wasa11,BCS17} and interpreted in Ref.~\cite{gal17} as due 
to a tightly bound $\Delta\Delta$ decay with reduced $\Delta \to N \pi$ phase 
space. The upper bump in (a) is consistent then with the d$^{\ast}(2380)\to 
\pi {\cal D}_{12}$ decay mode, in agreement with the $\pi d$ invariant-mass 
distribution shown in (b) that peaks slightly below the ${\cal D}_{12}$(2150) 
mass. 

\begin{figure}[!t]
\begin{center}
\includegraphics[width=0.48\textwidth,height=5.5cm]{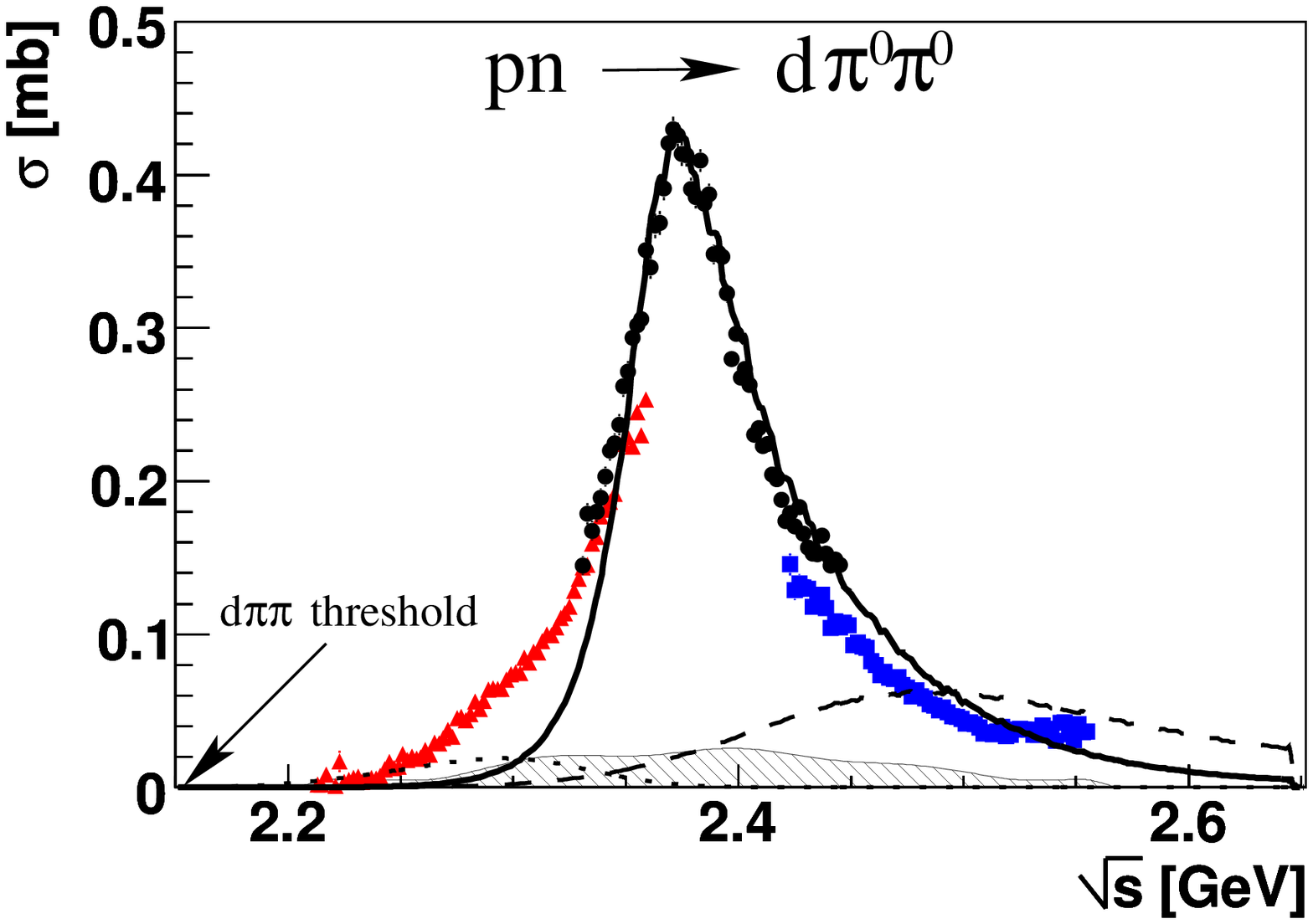}
\includegraphics[width=0.48\textwidth,height=5cm]{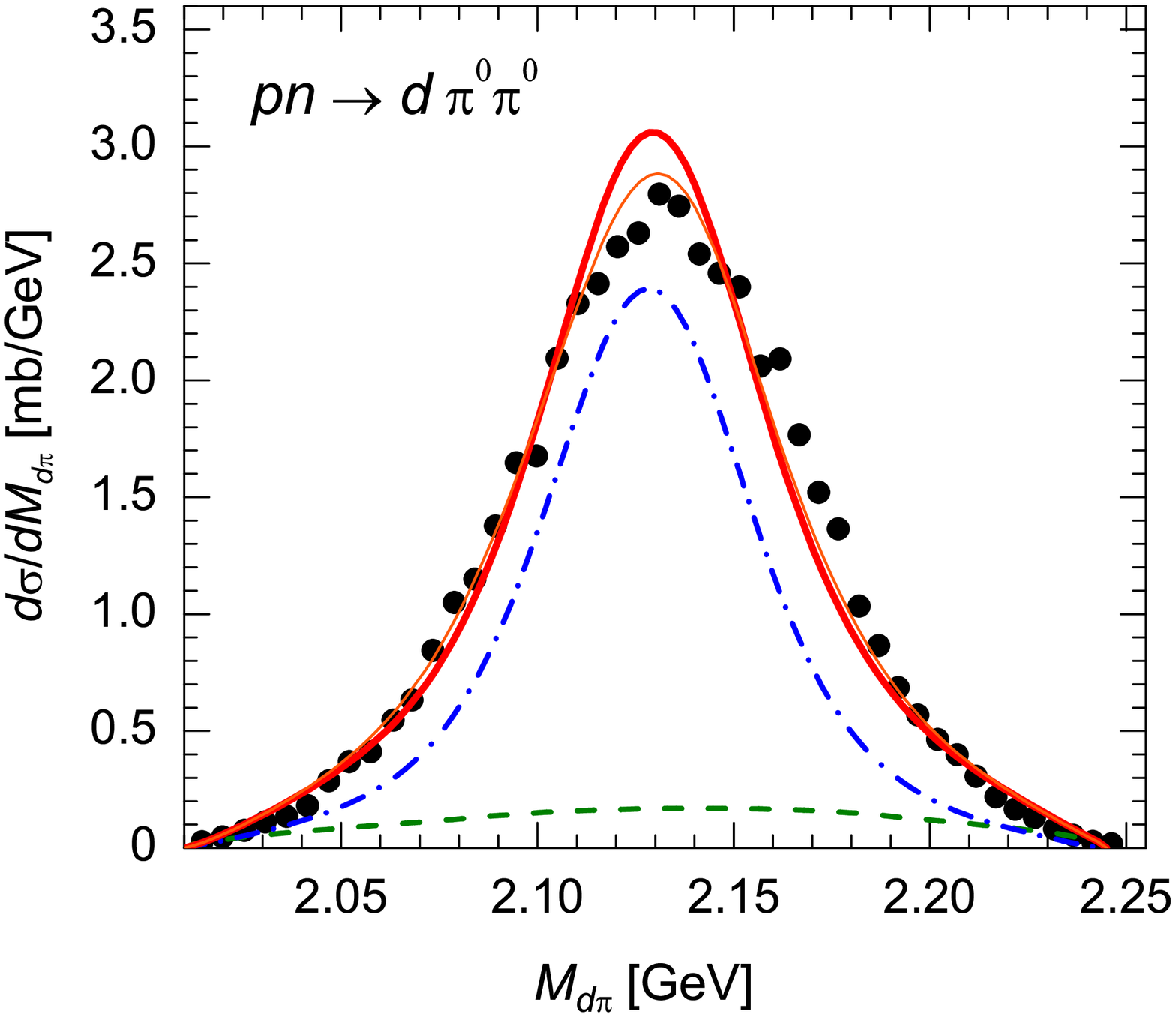}
\caption{The $pn\to d\pi^0\pi^0$ peak (left) and its $M_{d\pi}$ invariant-mass 
distribution (right) as observed by WASA-at-COSY \cite{wasa11}. The curves 
in the right panel are from Ref.~\cite{PK16}: the dot-dashed line gives the 
$\pi{\cal D}_{12}(2150)$ contribution to the two-body decay of the d$^{\ast}
$(2380) dibaryon, and the dashed line gives a $\sigma$-meson emission 
contribution. The solid lines are calculated $M_{d\pi}$ distributions 
for two input parametrizations of ${\cal D}_{12}(2150)$.} 
\label{fig:PK} 
\end{center} 
\end{figure} 

Theoretical support for the relevance of the ${\cal D}_{12}(2150)$ $N\Delta$
dibaryon to the physics of the d$^{\ast}$(2380) resonance is corroborated 
in Fig.~\ref{fig:PK} by showing in the right panel a $d\pi$ invariant-mass 
distribution peaking near the $N\Delta$ threshold as deduced from the 
$pn\to d\pi^0\pi^0$ reaction in which the d$^{\ast}$(2380) was discovered 
and which is shown for comparison in the left panel. However, the $M_{d\pi}$ 
peak is shifted to about 20 MeV below the mass of the ${\cal D}_{12}$(2150) 
and its width is smaller by about 40 MeV than the ${\cal D}_{12}$(2150) width, 
agreeing perhaps fortuitously with $\Gamma_{d^{\ast}(2380)}$. Both of these 
features, the peak downward shift and the smaller width, can be explained 
by the asymmetry between the two emitted $\pi^0$ mesons, only one of which 
is due to the $\Delta\to N\pi$ decay within the ${\cal D}_{12}$(2150).{
\footnote{I'm indebted to Heinz Clement for confirming this explanation.}} 

\begin{table}[hbt]
\begin{center}
\caption{d$^{\ast}$(2380) decay width branching ratios (BRs) calculated in 
Ref.~\cite{gal17}, for a total decay width $\Gamma_{d^{\ast}(2380)}$=75~MeV, 
are compared with BRs derived from experiment~\cite{BCS15,wasa17}.} 
\begin{tabular}{cccccccccc}
\hline
\% & $d\pi^0\pi^0$ & $d\pi^+\pi^-$ & $pn\pi^0\pi^0$ & 
$pn\pi^+\pi^-$ & $pp\pi^-\pi^0$ & $nn\pi^+\pi^0$ & $NN\pi$ & $NN$ & total \\ 
\hline 
BR(th.)  & 11.2 & 20.4 & 11.6 & 25.8 & 4.7 & 4.7 & 8.3 & 13.3 & 100 \\
BR(exp.) & 14$\pm$1 & 23$\pm$2 & 12$\pm$2 & 30$\pm$5 & 6$\pm$1 & 6$\pm$1 & 
$\leq$9 & 
12$\pm$3 & 103 \\ 
\hline
\end{tabular}
\label{tab:BR}
\end{center}
\end{table}

Recalling the $\Delta\Delta$ -- $\pi{\cal D}_{12}$ coupled channel nature of 
the d$^{\ast}$(2380) in our hadronic model~\cite{GG13,GG14}, one may describe 
satisfactorily the d$^{\ast}$(2380) total and partial decay widths in terms 
of an incoherent mixture of these relatively short-ranged ($\Delta\Delta$) 
and long-ranged ($\pi{\cal D}_{12}$) channels. This is demonstrated in 
Table~\ref{tab:BR} where weights of $\frac{5}{7}$ and $\frac{2}{7}$ for 
$\Delta\Delta$ and $\pi{\cal D}_{12}$, respectively, are assigned to an 
assumed value of $\Gamma_{d^{\ast}\to NN\pi\pi}$=60~MeV~\cite{gal17}. This 
choice yields a branching ratio for $\Gamma_{d^{\ast}\to NN\pi}$ which does 
not exceed the upper limit of BR$\leq$9\% determined recently from {\it not} 
observing the single-pion decay branch~\cite{wasa17}. A pure $\Delta\Delta$ 
description leads, as expected, to BR$\ll$1\%~\cite{dong17}.

\section{Hexaquark, diquark and you-name-it-quark models for d$^{\ast}$(2380)} 
\label{sec:disc} 

In this concluding section we comment on the applicabilty of several 
quark-based models to the d$^{\ast}$(2380) dibaryon resonance. Interestingly, 
the same main Beijing group arguing for a compact hexaquark structure of 
the d$^{\ast}$(2380) has voiced recently~\cite{dong19b} reservations on 
the ability of their underlying model to reach the observed level of the ELPH 
measured cross section $\sigma(\gamma d\to d^{\ast}(2380)\to d\pi^0\pi^0)$ 
shown in Fig.~\ref{fig:fix}. The hexaquark cross section calculation 
underestimates the BW contribution in the figure, $\approx$18~nb at 
the nominal resonance energy, by about a factor of 20~\cite{dong19b}. 

Another quark-based model suggestion was made recently by a faction of the 
Beijing group~\cite{shi19}. These authors tried to fit the d$^{\ast}$(2380) 
within a diquark ($\cal D$) model in terms of a bound system of three vector 
diquarks. However, Gal and Karliner~\cite{gk19} noted that the $I=0,J^P=1^+$ 
deuteron-like and the $I=1,J^P=0^+$ virtual-like 3$\cal D$ states in the 
particular diquark model considered are located about 200-250 MeV above the 
physical deuteron, where no hint of irregularities in the corresponding $NN$ 
phase shifts analyses occur. In fact no resonance feature in the corresponding 
partial-wave phase shifts up to at least $W=2.4$~GeV has ever been reliably 
established~\cite{SAID}. Moreover, it was shown by these authors~\cite{gk19} 
that if the d$^{\ast}$(2380) structure were dominated by a 3$\cal D$ 
component, its decay width would have been suppressed by at least an 
isospin-color recoupling factor 1/9 with respect to the naive $\Delta\Delta$ 
hadronic estimate of 160~MeV width, bringing it cosiderably below the deduced 
value of $\Gamma_{d^{\ast}(2380}\approx 70$~MeV. 

We end with a brief discussion of possible 6q admixtures in the essentially 
hadronic wavefunction of the d$^{\ast}$(2380) dibaryon resonance. For this 
we refer to the recent 6q non-strange dibaryon variational calculation in 
Ref.~\cite{PPL15} which depending on the assumed confinement potential 
generates a $^3S_1$ 6q dibaryon about 550 to 700~MeV above the deuteron, 
and a $^7S_3$ 6q dibaryon about 230 to 350~MeV above the d$^{\ast}$(2380). 
Taking a typical 20~MeV potential matrix element from deuteron structure 
calculations and 600~MeV for the energy separation between the deuteron and 
the $^3S_1$ 6q dibaryon, one finds admixture amplitude of order 0.03 and 
hence 6q admixture probability of order 0.001 which is compatible with that 
discussed recently by Miller~\cite{miller14}. Using the same 20~MeV potential 
matrix element for the $\Delta\Delta$ dibaryon candidate and 300~MeV for the 
energy separation between the d$^{\ast}$(2380) and the $^7S_3$ 6q dibaryon, 
one finds twice as large admixture amplitude and hence four times larger 6q 
admixture probability in the d$^{\ast}$(2380), altogether smaller than 1\%. 
These order-of-magnitude estimates demonstrate that long-range hadronic and 
short-range quark degrees of freedom hardly mix also for $\Delta\Delta$ 
configurations, and that the d$^{\ast}$(2380) is extremely far from a pure 6q 
configuration. This conclusion is at odds with the conjecture made recently 
by Bashkanov, Brodsky and Clement~\cite{BBC13} that 6q CC components dominate 
the wavefunctions of the $\Delta\Delta$ dibaryon candidates ${\cal D}_{03}$, 
identified with the observed d$^{\ast}$(2380), and ${\cal D}_{30}$. 
Unfortunately, most of the quark-based calculations discussed in the present 
work combine quark-model input with hadronic-exchange model input in a loose 
way~\cite{lu17} which discards their predictive power.


\end{document}